# Crosstalk Impacts on Homogeneous Weakly-Coupled Multicore Fiber Based IM/DD System

Lin Gan,[1] Jiajun Zhou,[1] Liang Huo,[1] Li Shen,[1] Chen Yang,[2] Weijun Tong,[2] Songnian Fu,[1] Ming Tang,[1,*] and Deming Liu[1]

[1]Wuhan National Lab for Optoelectronics (WNLO) & National Engineering Laboratory for Next Generation Internet Access System, School of Optics and Electronic Information, Huazhong University of Science and Technology

[2]State Key Laboratory of Optical Fiber and Cable Manufacture Technology, Yangtze Optical Fiber and Cable Joint Stock Limited Company (YOFC). R&D Center

*tangming@mail.hust.edu.cn

*Abstract*—We numerically discussed crosstalk impacts on homogeneous weakly-coupled multicore fiber based intensity-modulation/direct-detection (IM/DD) systems taking into account mean crosstalk power fluctuation, walk-off between cores, laser frequency offset, and laser linewidth.

## I. Introduction

As a promising realization of space-division multiplexing (SDM) technology, weakly-coupled multicore fiber (WC-MCF) has been widely discussed in various scenarios, such as inter-datacenter interconnect (DCI), passive optical network (PON), front-haul and long-haul transmission [1-3]. However, nowadays there is no unified standardization about WC-MCF. One of the most critical issues is to determine appropriate WC-MCF's fiber parameters, such as core diameter, difference of refractive index, core pitch. Because these fiber parameters will obviously affect the characteristics of inter-core crosstalk (XT) in time-domain and frequency-domain [4, 5]. And the IC-XT will further affect the transmission performance. Therefore, it is necessary to figure out the XT penalty for each transmission scenario.

To the best of our knowledge, most of investigations about XT penalty and its fluctuation are discussed experimentally, which will greatly limit the scope and depth of analysis [4,6]. In this manuscript, we will focus on WC-MCF based intensity-modulation/direct-detection (IMDD) systems. Taking the pulse amplitude modulation (PAM) signals as an example, we will demonstrate some issues that have not been studied yet.

Our key findings include:
a) The fluctuation of signal-to-noise ratio (SNR) induced by large XT will significantly decrease from around 20 dB to 1 dB with the increasing of walk-off between cores.
b) The impacts of XT will be similar to additive white Gaussian white noise (AWGN) with large walk-off or higher modulation format.
c) The frequency offset between different lasers will lead to serious carrier beating noise. Unless the frequency offset is less than 100 kHz, the maximum acceptable XT power will decrease from -30 dB to -40 dB.

Therefore, for short-reach IM/DD system, heterogeneous WC-MCF should be better than homogeneous WC-MCF due to its large walk-off. The laser frequency offset among different cores should be as small as possible to avoid extra carrier beating noise. Under this assumption, the XT impacts on transmission signals could be modeled as additive white Gaussian noise (AWGN) to support MCF design.

## II. Simulation setup

Fig.1 is a typical WC-MCF based short-reach IM/DD system. In each core, two independent 28 Gbaud PAM-2/PAM-4 symbols are generated from a pseudorandom bit sequence of word length $2^{15}-1$ (PRBS-15) before being up-sampled 16 times and filtered with a root raised cosine filter of 0.18 roll-off factor. The PAM-2/PAM-4 signals are modulated by Mach-Zehnder modulator (MZM) with arbitrary waveform generators (AWG). Similar to the scenario of combining vertical cavity surface emitting laser (VCSEL) array with MCF, the optical carrier's frequency in each channel may be different with each other, which is caused by unavoidable fabrication error. We set optical signal-to-noise ratio (OSNR) by adding AWGN to simulate noises in real situations with output average power of 0 dBm. After transmitted over 3.0-km MCF link, the signals in each core are directly detected by a photodiode (PD). After sampled by a real-time digital storage oscilloscope (DSO), offline digital signal processing (DSP) is conducted, which consists of a low-pass filter, a maximum variance timing recovery, and a T/2 symbol-spaced decision-directed least mean square (DD-LMS) algorithm with 21 feed-forward taps (FFT) and 22 feedback taps (FBT).

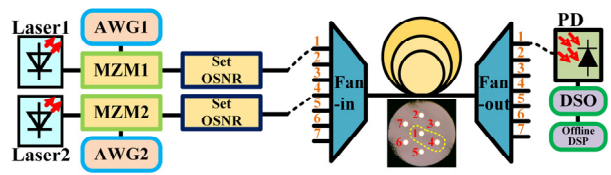

Fig. 1. *Simulation setup for MCF based short reach IM/DD system*

Fig.2 shows the proposed channel model for homogeneous WC-MCF to support simulation of the propagation effects in a MCF taking into account attenuation, polarization mode dispersion (PMD), walk-off, chromatic dispersion (CD), dispersion slope, XT and nonlinear Kerr effects [7]. We use the split-step Fourier method to solve MCF with XT described by

National Natural Science Foundation of China under Grant (61722108, 61331010, 61205063, 61290311); Major Program of the Technical Innovation of Hubei Province of China (2016AAA014); the Open Fund of State Key Laboratory of Optical Fiber and Cable Manufacture Technology, YOFC (SKLD1706) and the Fundamental Research Funds for the Central Universities': HUST: 2018KFYXKJC023.

coupling matrix in each calculation step size. In addition, it should be noted that the mean XT power is set manually.

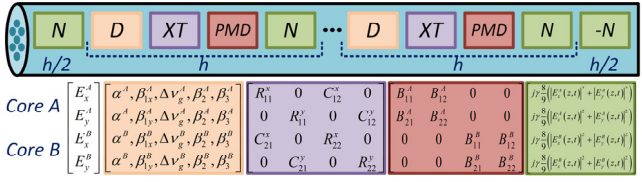

Fig. 2. *Simulation model for homogeneous WC-MCF*

The walk-off, center frequency and linewidth of two lasers, is 0.1 ps/m, 193.12 THz, and 50 kHz, respectively. Fig.3(a) and 3(b) show the eye-diagrams of the received PAM-2/PAM-4 signals after PD without XT, respectively. Fig.4(a) and 4(b) show the corresponding signals of PAM-2/PAM-4, respectively, after equalizing with SNR of 45 dB.

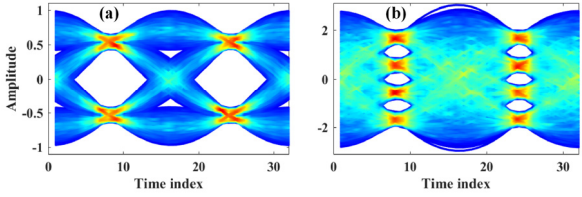

Fig. 3. *Eyediagram of PAM2 (a) and PAM4 (b)*

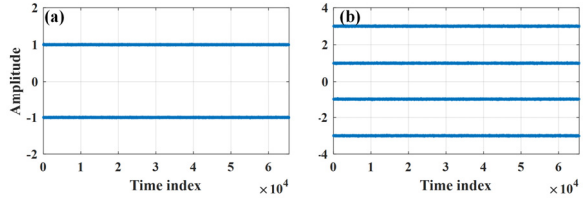

Fig. 4. *Signals after DD-LMS equalizing of PAM2 (a) and PAM4 (b).*

### III. SIMULATION RESULTS

Fig.5(a) shows the XT impacts on PAM-2/PAM-4 signals with three different SNR level including 45 dB, 35 dB, and 25 dB. All the simulations repeated 100 times. It can be observed that the degradation of high SNR level under the same XT power is significantly larger than that of low SNR level. Fig.5(b) shows the $R^2$ of Gaussian fitting of the recovered signals. It can be founded that not all the $R^2$ are close to 1.0 which means the XT impacts on signals cannot be treated as AWGN in some cases.

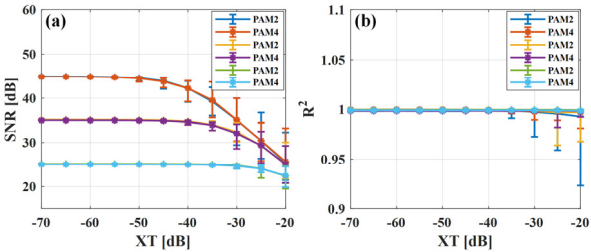

Fig. 5. *Transmission performance with different XT power. (a) the SNR of signals after equalization, (b) the $R^2$ of Gaussian fitting.*

Due to the fact that the frequency-dependent characteristics of XT are dominated by walk-off, Fig.6(a) shows the XT impacts on PAM-2/PAM-4 signals with different walk-off. The SNR level is 45 dB. XT has little impact on the SNR when XT is -50 dB/3 km. However, when XT is -25 dB/3 km, the average SNR is decreased to 30 dB. In addition, the fluctuation of SNR is more than 20 dB when walk-off is less than 0.01 ps/m. Namely, the XT impacts on transmission signals will be more stable (about 1dB) with large walk-off of 1.0 ps/m. Fig.6(b) shows the results of Gaussian fitting. It can be concluded that the impacts of XT will be more similar to AWGN with large walk-off or higher modulation format.

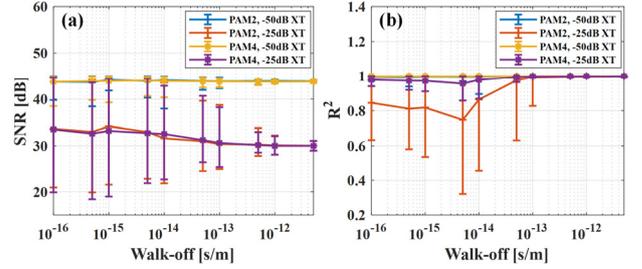

Fig. 6. *Transmission performance with different walk-off. (a) the SNR of signals after equalization, (b) the $R^2$ of Gaussian fitting.*

Fig.7(a) and 7(b) demonstrate one realization of the distribution of PAM-2 signals with walk-off and XT power of $10^{-4}$ ps/m and -25 dB/3 km, respectively. Due to small walk-off, the de-correlation bandwidth of XT will be large. Therefore, the XT impacts can be treated as signal copies with two peaks for each level of PAM-2. Fig.7(c) and 7(d) demonstrate one realization of the distribution of PAM-2 with walk-off and XT power of 1.0 ps/m and -25 dB/3 km. Due to small de-correlation bandwidth of XT, the impacts of XT can be treated as AWGN.

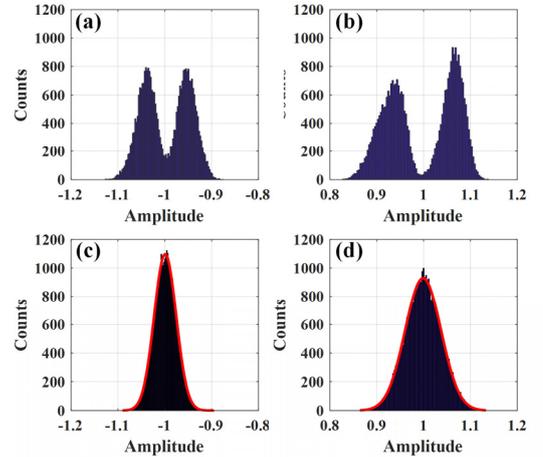

Fig. 7. *Signal distibutions after equalization for PAM2 with XT power -25 dB/3 km. The walk-off is $10^{-16}$ s/m for (a) and (b). And the walk-off is $10^{-12}$ s/m for (c) and (d).*

In order to find out the nature of the impact of walk-off on XT, Fig.8 shows the actual XT power with different walk-off even if the target XT power is -50 dB/3 km. Each situation repeated 1000 times to get the statistical characteristics of XT power. However, the real XT power fluctuates obviously with small walk-off, such as 0.01 ps/m or less, which causes the SNR fluctuation of Fig.5(a) and Fig.6(a). The reason why the real XT power fluctuates obviously is that the XT power of different optical frequencies have strong correlation. Further, it means, for long-term time-dependent XT, heterogeneous WC-MCF based links are more stable with time than homogeneous WC-MCF based links due to its large walk-off. XT of different optical frequencies can be treated independent with each other. Therefore, the mean XT power of all frequencies will be more stable.

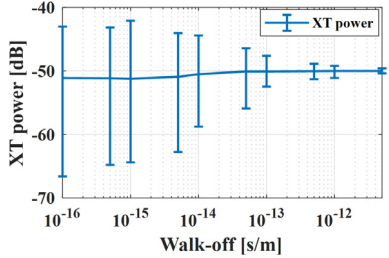

Fig. 8. *XT power fluctuation with different walk-off.*

In practical situation, due to unavoidable fabrication error, the center frequencies of the two lasers are usually different. Therefore, Fig.9(a) shows the XT impacts with different frequency offset of the two lasers taking into account with walk-off of 0.1 ps/m. Each situation repeated 100 times. Under the same XT power, the average SNR will be obviously decreased when frequency offset is larger than 100 kHz. Fig.9(b) shows the waveform received by PD with frequency offset and target XT power of 5.0 MHz and -25 dB, respectively. There is strong beating noise of optical carriers with period about 0.2 μs.

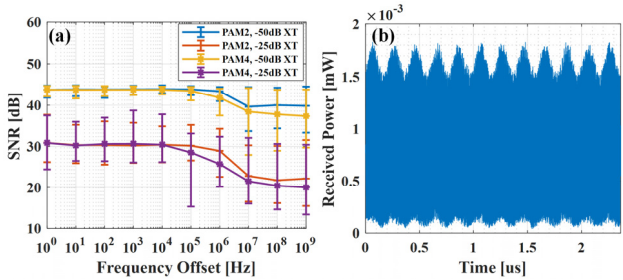

Fig. 9. *(a) Transmission performance with different laser frequency offset. (b) The receive waveform after photodetector with frequency offset of 5.0 MHz.*

Fig.10 shows the XT impacts of different XT power when 100 MHz frequency offset is added. Compared with Fig.5(a), it can be observed that the maximum acceptable XT power (with 0.1 dB SNR penalty) will decrease from -30 dB to -40 dB when SNR level of received signals is 25 dB which is more similar to the real SNR in practical situation.

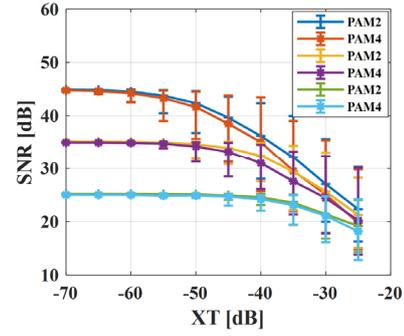

Fig. 10. *Transmission performance with different XT power under laser frequency offset of 100 MHz.*

Lastly, laser linewidth is also an important property for transmission systems, even if it is a IM/DD system. Therefore, Fig.11 shows the XT impacts with different linewidth. It can be concluded that XT is independent with laser linewidth.

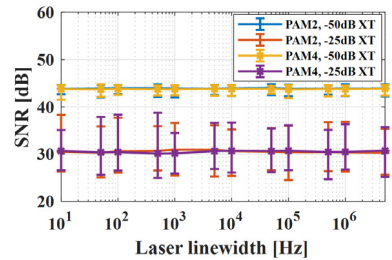

Fig. 11. *Transmission performance with different laser linewidth.*

IV. CONCLUSIONS

We have studied the XT impacts on the PAM-2/PAM-4 signals for short-reach IM/DD system with mean crosstalk power fluctuation, walk-off between cores, laser frequency offset, and laser linewidth taking into account. It can be concluded that, for short-reach IM/DD system, heterogeneous WC-MCF should be better than homogeneous WC-MCF due to its large walk-off. The frequency offset between lasers will lead to serious carrier beating noise unless it is less than 100 kHz. Only in this situation, the XT impacts on transmission signals could be modeled as AWGN to support MCF link design.